\documentstyle[sprocl,epsfig]{article}

\def\be{\begin{equation}}
\def\ee{\end{equation}}
\def\bea{\begin{eqnarray}}
\def\eea{\end{eqnarray}}

\begin{document}
\title{Gamma-ray bursts and gravitational radiation from black hole-torus systems}
\author{M.H.P.M. van Putten}
\address{MIT 2-378, Cambridge, MA 02139, USA\\E-mail: mvp@schauder.mit.edu}

\maketitle\abstracts{ 
Cosmological gamma-ray bursts (GRBs) are probably powered by
systems harboring a rotating black hole. This may result from
hypernovae or black hole-neutron star coalescence. We identify
short/long bursts with hyper- and suspended-accretion states 
around slowly/rapidly rotating black holes
 [van Putten \& Ostriker, ApJL, 552, L31, 2001]. 
Black holes may be activated into producing outflows by a surrounding 
torus magnetosphere, in the form of baryon poor jets as input to 
the observed GRB/afterglow emissions. Here, we attribute these 
outflows to a differentially rotating gap in an open flux-tube 
along the axis of rotation of the black 
hole [van Putten, Phys. Rep., 345, 1 (2001)].
A high incidence of the black hole luminosity into the surrounding
matter is expected by equivalence in poloidal topology to pulsar
magnetospheres. For long bursts, this predicts a large fraction
of the black hole spin-energy emitted in gravitational
waves by a quadrupole moment in the surrounding torus
 [van Putten \& Levinson, ApJL, 555, L41, 2001].
This suggests that long GRBs may be the most powerful
LIGO/VIRGO burst sources of gravitational waves in the 
Universe with an expected duration of 10-15s on a horizontal
branch of 1-2kHz in the $\dot{f}(f)-$diagram
 [van Putten, Phys. Rev. Lett., 87, 091101, 2001].
Gravitational wave-emissions from GRBs, therefore, promise 
calorimetric evidence for Kerr black holes.
}

\section{Introduction}

Cosmological gamma-ray bursts (GRBs) are the most enigmatic
events in the Universe (Fig. 1). Their emissions are characteristically
nonthermal in the few hundred keV range
with a bi-modal distribution in durations, of short bursts 
around 0.3s and long bursts around 30s.\cite{kou93}
Redshift determinations from long bursts indicate a
cosmological origin, probably associated with the formation
of young massive stars.\cite{pac98,blo00} GRBs, therefore, are probably most
frequent within a redshift $z=1-2.$ The recent proposal \cite{bro00} that
GRB relics could be found in some of the galactic
soft X-ray transients (SXTs), notably so GRO J1655-40 \cite{isr99} 
and V4641 Sgr,\cite{oro01} further suggests an association with black 
holes of about $3-14M_\odot$.

The inner engine producing the GRBs should be
energetic and compact. Angular momentum forms a canonical energy reservoir,
and GRB inner engines are probably no exception. This is consistent
with breaking of spherical symmetry.\cite{woo93} Hypernovae, then, follow up 
on this idea by postulating massive young stars in binaries as their 
progenitors. High angular momentum is also present in coalescing black 
hole-neutron star systems.

\begin{figure}
\center{\epsfig{file=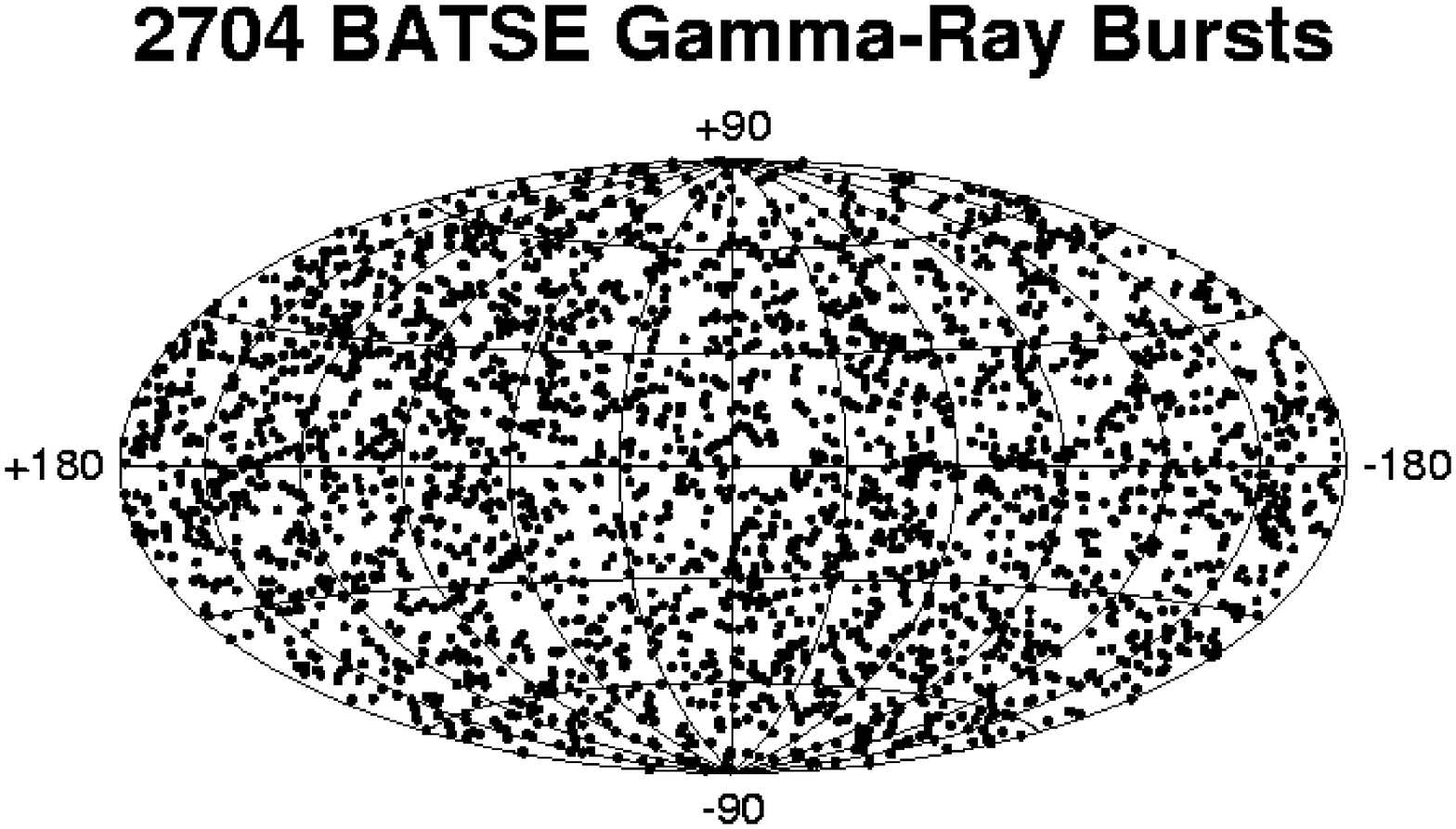,height=80mm,width=110mm}}
\caption{Shown are the locations of 2704 GRBs from the GRBs from
         the BATSE Catalogue over a nine-year period. The projection is
		 in galactic coordinates. (Courtesy of NASA Marshall Space Flight
		 Center, Space Sciences Laboratory.) The isotropic distribution
		 and a $<V/V_{max}>$ distinctly less than the Euclidean value of
		 1/2 (Schmidt 1999)
		 are indicative for their cosmological origin. For long bursts,
		 this is further evidenced by redshifts of order unity from GRB
		 afterglow emissions and a probably association with star-forming
		 regions (Paczynski 1998; Bloom et al. 2000).
		 Hence, long bursts are expected to track closely the
		 star-formation rate, which peaks at about $z=1-2$.
		 Remnants of GRBs are potentially found in some of the  
		 Soft X-ray transients, with notable candidates
		 GRO J1655-40 (Israelian et al, 1999) and V4641Sgr
		 (Orosz et al., 2001)}
\end{figure}
Here, I discuss some physical aspects of black hole-torus systems 
and a few observational predictions. Black holes may be active in
accord with the Rayleigh criterion, in contact with a torus
magnetosphere supported by baryonic matter.
This is expected to be manifest in non-thermal emissions, in  
outflows along an open flux-tube on the axis of rotation of the
black hole and various emissions from a surrounding torus:

\begin{itemize}
\item 
At supercritical magnetic field-strengths, outflows may spontaneously be 
created by frame dragging-induced potential differences, 
as follows from a perturbative calculation in Hawking's approach.\cite{mvp00}
In competition with equilibration by charge separation, pair-creation is 
expected to be confined to a gap along the open flux-tube. Asymptotic boundary 
conditions on the horizon and infinity, linked by current continuity, define 
the net dissipation in the gap, and hence a powerful outflow whenever the black 
hole spins more rapidly than twice the angular velocity of the torus.\cite{mvp01b} 
\item 
The torus is expected to be luminous in several channels, notably
so in gravitational radiation, neutrino emissions and Poynting flux winds.
Powered by the spin-energy of the black hole \cite{mvp01a}, the fluence
in gravitational waves may reach about 1\% of the black hole mass.  
The coupling to the rotational energy of the black hole operates
by equivalence in poloidal topology to pulsar magnetospheres.\cite{mvp99,bro01} 
Detection by LIGO/VIRGO of these emissions may result
in calorimetric evidence of Kerr black holes \cite{mvp01d}, whenever
the fluence is determined to be in excess of the rotational energy of
a rapidly rotating neutron star.
\end{itemize}

These prospects suggest several observational predictions.

Increasing evidence towards clustering in the true energy in GRB emissions 
\cite{fra01,pan01,pir01} indicates a standard opening angle of about 
$\theta_H\simeq 35^o$ of the open magnetic flux-tube on the horizon of the 
black hole \cite{mvp01c}.
This predicts a cut-off $\theta_j\le 35^o$ in the observed opening angles 
$\theta_j$ on the celestial sphere in any large statistical sample.
Furthermore, with outflows created in long and short bursts alike,
HETE-II may detect afterglows also to short GRBs 
\cite{mvp01a} (see also \cite{nar01}). 
The latter is expected to differ mostly in net fluence, as their
inner engine operates for a significantly shorter time than in long GRBs.

Gravitational radiation appears to be a major channel in the emissions
from the torus in long bursts, along with emissions in neutrinos and
winds coming off the torus. Calculations in the
suspended-accretion state indicate a net luminosity in gravitational
waves of about one-third the net luminosity of the black hole.\cite{mvp01d}
This amounts to a fluence $E_{gw}$ about $1\%$ of the mass-energy of 
the central black hole. The frequency $f_{gw}$ in
these emissions at twice the Keplerian frequency of the torus, as it
develops a quadrupole moment in its mass distribution, is expected to be 
about 1-2kHz for a black hole mass of about $10M_\odot$.
This range overlaps with the design bandwidth of 0.1-1.5kHz of 
the upcoming Laser Interferometric Gravitational Wave Observatories
LIGO/VIRGO.

This raises an unanticipated prospect: calorimetric evidence for Kerr black
holes from the emission in gravitational waves from the torus.\cite{mvp01e} 
Indeed, consider the product $\alpha=2\pi E_{gw}f_{gw}$, which expresses a
measure for the ratio of rotational energy to the linear size of the inner
engine. It appears that $\alpha$ from black hole-torus systems may reach 
values in excess of those attainable by rapidly rotating neutron stars.
A LIGO/VIRGO detection of a large $\alpha$, therefore, individually or as 
an average over a sample of detections, would be evidence for the Kerr 
relationship $E_{rot}\sim M/3$ between the rotational energy $E_{rot}$ and 
the mass $M$ of a rapidly rotating black hole.

The proposed association of gamma-ray bursts to black hole-torus
systems will be reviewed in \S2, and prospects for
GRBs as potential LIGO/VIRGO sources is outlines in \S3. We close
with a comment on the potential for calorimetric evidence of
Kerr black holes in GRBs.

\section{A theory of gamma-ray bursts from black hole-torus systems}

A black hole-torus system is of compact dimension, consistent with
the short time-variability in the GRB light-curves
and the proposed GRB-SXT association.
The mass in the surrounding torus or disk will be limited, 
in both the hypernovae and binary black hole-neutron star coalescence
scenario. This introduces relatively short time-scales of accretion,
leaving a central Kerr black hole as the major energy reservoir.
This poses two questions: what accounts for the duration in
long GRBs and how can the rotational energy of the black hole
create baryon poor jets?

\subsection{Formation of black hole-disk or torus systems}

A black hole-torus system may form from binary black hole-neutron
star coalescence. Here, the neutron star gradually approaches the
black hole by angular momentum loss in gravitational radiation.
The neutron star will then be subject to tidal
interactions, which may lead to break-up outside the
inner most stable circular orbit (ISCO) when the central
black hole is sufficiently small in mass. For this to happen,
canonical estimates provide a bound of $3.7M_\odot$ on 
non-rotating black holes and a bound of $28M_\odot$ on
rapidly rotating black holes. This indicates a substantially
wider window of mass for the rotating case. It follows that
a torus is more likely to form around a Kerr black hole.\cite{pac91}

The collapsar, failed supernova or hypernova scenario
envisions the collapse of the center of a young massive
star with high angular momentum. The origin of the angular
momentum is most likely orbital angular momentum from the
progenitor binary system. While the details of 
orbital angular momentum transfer into the collapsing star are
somewhat uncertain, collapse of a rapidly rotating object
is expected to result in a compact core surrounded by matter
stalled against an angular momentum barrier. If the core
forms a black hole in prompt collapse then, furthermore,
the black hole will have a {\em minimum} mass,
sufficient to account for the angular momentum $J_H$ in view of
the Kerr constraint $J_H^2\le M^2$ (in geometrical units,
with $M$ denoting the Schwarzschild radius $Gm/c^2$,
where $G$ is Newton's constant, $m$ the mass of the black hole 
and $c$ the velocity of light) \cite{mvp01a}. For example,
a Lane-Emden relationship with polytropic index $n=3$
for the progenitor star
gives $M\ge 10M_\odot$, consistent with the observed
range of $3-14M_\odot$ in SXTs shown in Fig. 2.

\begin{figure}
\center{\epsfig{file=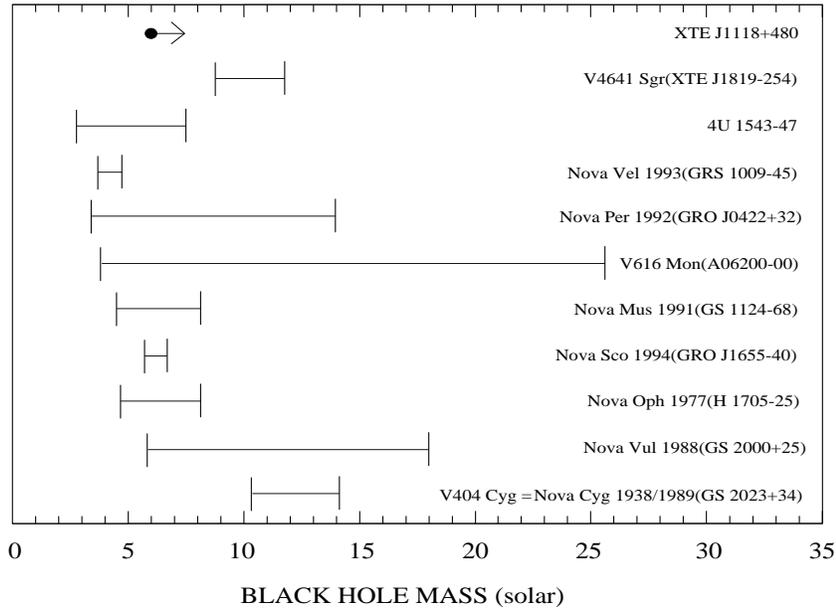,height=80mm,width=110mm}}
\caption{Shown is the distribution of black hole masses in X-ray novae.
	 The top four are XTE J118+408 (McClintock et al., 2001),
	 V4641 Sgr (Orosz et al., 2001), 4U 1543-47 (Orosz et al., 1998)
	 and Nova Vel 1993 (Filippenko et al., 1999); the lower seven
	 are from Bailyin et al. (1998). This mass distribution 
	 manisfests a certain diversity in black hole masses of about
	 $3-14M_\odot$. [Reprinted from van Putten, {\em Physics Reports},
	 345 \copyright 2001 Elsevier B.V.]
}
\end{figure}

In both scenarios, a magnetized neutron star or young massive
star - represented by a magnetic moment density -- will result
in a disk or torus endowed with a net poloidal flux.
It may be appreciated that the mass in the disk or torus
a finite in a much more stricter sense than in analogous
configurations believed to exist in active galactic nuclei.
With magnetic regulated accretion, accretion of $0.1M_\odot$ 
becomes fairly rapid on a time-scale of a second or less onto
onto a $10M_\odot$ mass black hole. Depleting the surroundings 
of baryonic matter inevitably prevents any further activity of 
the black hole. This suggests that additional physical processes
should account for the relatively long duration in long bursts.
In a recent proposal, the magnetic moment density of the surrounding
torus is believed to permit a suspended-accretion state for the 
duration of spin-down of the central black hole. A bi-modal
distribution of durations then occurs when the ratio of black 
hole-to-disk or torus mass is large.\cite{mvp01a}

\subsection{The lowest energy state of the black hole}

The black hole will be surrounded by a torus magnetosphere,
supported by the accretion disk.
The black hole will adjust to a lowest energy state by
developing an equilibrium magnetic moment\cite{mvp01b}
\begin{eqnarray}
\mu_H^e\simeq aBJ_H, 
\end{eqnarray}
where $a=J_H/M$ denotes the
specific angular momentum of a black hole with mass
$M$ and angular momentum $J_H$ and $B$ denotes the
(average) poloidal magnetic field. This
results from a minimum in the energy
${\cal E}(q)=(1/2)Cq^2-\mu_HB$, where $C\simeq 1/r_H$
denotes the capacitance of the black hole, $q$ the
charge on the horizon, and Carter's identity \cite{car68}
$\mu_H=qJ_H/M$ (``no fourth hair"). The minimum of
${\cal E}$ at $q\simeq BJ_H$ recovers Wald's result.\cite{wal74} 
Similar results are found in a largely force-free magnetosphere.\cite{lee01}
This equilibrium magnetic moment preserves an essentially maximal 
and uniform horizon flux. This serves to preserve a strong coupling 
to the magnetosphere and, hence, to the inner face of the 
surrounding torus. It also permits the black hole
to support
an open flux-tube to infinity along its
axis of rotation, particularly so in a suspended accretion state. 
Frame-dragging will act on this flux-tube, to
to produce a differentially rotating gap for the creation
of baryon poor outflows.

\subsection{Hyper- and suspended-accretion states for a ring}

GRBs show a bi-modal distribution in durations, as show in
Fig. 3. We attribute this to different modes of angular
momentum losses in black hole plus disk or torus systems.

As magnetic fields can be very efficient in mediating
angular momentum transport, magnetic regulated
accretion times tend to be short. Indeed, the
accretion of a magnetized ring is illustrative, showing
evolution towards a finite-time singularity\cite{mvp01a}
\begin{eqnarray}
\varpi=(1-t/t_f)^p
\end{eqnarray}
of its radius $\varpi=R(t)/R(0)$, where $R(0)$ denotes the initial
radius and $t_f$ the final time of collapse. Here, $p=1/2$ and $2$ for
a split monopole geometry (SMG) and toroidal field geometry
(TFG), respectively. The final time in both field geometries
is governed by the ratio of kinetic-to-magnetic energy 
$\delta E_k/\delta E_B$ - a free parameter, at present not well constrained
from first principles or current numerical simulations. For a
fiducial value of 10$^2$, SMG applied to the initial evolution
ring and TFG applied to the final evolution of the ring
indicates accretion times $t_f$ of a few seconds or less -- consistent
with the time-scale of short bursts. This analysis does not
take into account any action by the black hole back back onto
the ring, i.e.: the results apply to the accretion 
onto slowly rotating or Schwarzschild black holes.
\begin{figure}
\center{\epsfig{file=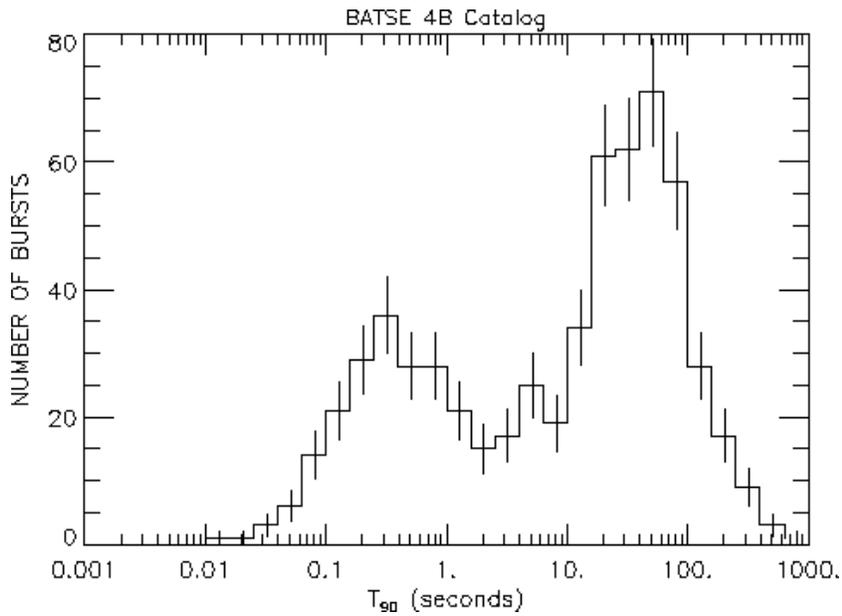,height=80mm,width=110mm}}
\caption{Shown is the bi-modal distribution of GRB durations of the
         4B Catague, set by a T90 duration parameter, on lightcurves
		 integrated over all 4 channels ($E>$20keV). (Courtesy of
		 NASA Marshall Space Flight Center, Space Sciences Laboratory.)
		 The population of long bursts is probably associated with 
		 young massive stars and, hence, with a redshift of $z=1-2$.
		 This indicates a redshift corrected mean value of the
		 intrinsic duration of about 10-15s. 
		 In van Putten \& Ostriker (2001), shorts bursts are identified with
		 magnetic regulated hyperaccretion onto slowly rotating black 
		 holes, and long bursts with rapidly rotating black holes in
		 a state of suspended accretion. Long bursts are potential
		 LIGO/VIRGO sources of gravitational radiation by gravitational
		 radiation from the torus, derived from the spin-energy of the
		 black hole. This indicates a mean duration of 10-15s of
		 gravitational wave-emissions commensurate with the redshift corrected 
		 GRB-event, for a cosmologically nearby sample within the detection
		 sensitivity of LIGO/VIRGO.}
\end{figure}

What then, may account for the duration of long bursts?
A ring surrounding a rapidly rotating black hole
will support, by a magnetic moment density, poloidal magnetic 
field-lines connected the ring to the horizon. 
These field-lines permit energy and angular momentum transport 
from the spin of the black hole into the ring. This process operates
by equivalence in poloidal topology to pulsar magnetospheres,
wherein the ring and the horizon of the black hole are, respectively, 
equivalent to a pulsar and infinity. The angular velocity of the 
equivalent pulsar
is the relative angular velocity between the ring and the black hole.
When the black hole spins more rapidly than the ring, the ring receives
angular momentum {\em like a pulsar being spun-up when
infinity wraps around it}.

The ring -- an element of the disk or torus -- will assume
a suspended-accretion state arises when radiative losses in
energy and angular momentum to infinity are replenished by
gain from the central black hole. For a balance on magnetic
torques alone, we find a critical angular velocity 
$\Omega_H^*$ of the black hole: \cite{mvp01a}
\begin{eqnarray}
\Omega_H^*=\Omega_T[1+(f_w/f_H)^2|\ln(\theta_0/2)|^{-1}]
\end{eqnarray}
for fractions $f_w$ and $f_H$ of magnetic flux
which connect to infinity (in an outgoing
Poynting flux-wind) and the black hole (in an ingoing Poynting flux-wind), 
respectively, where $\theta_0$ denotes the minimum poloidal angle in TFG.
This state lives as long as the black hole spins rapidly \cite{mvp01a}
\begin{eqnarray}
t_{long}=88s \left(\frac{M}{10M_\odot}\right)
              \left(\frac{M/M_d}{100}\right)
			  \left(\frac{E_k/E_B}{100}\right)
              g^2(\theta_0),
\end{eqnarray}
where $g(\theta_0)$ denotes a geometrical factor of order unity.
We conclude that
{a bi-modal distribution in duration occurs due to hyper- and
suspended-accretion states whenever the ratio $M/M_d$ is large}.

\subsection{Outflows from a differentially rotating gap} 

Frame-dragging introduces differential rotation along the axis of
rotation of a Kerr black hole. Flux-tubes on a differentially
rotating space-time background thus tend to develop potential differences 
by Faraday-induction. These potentials may drive large-scale
currents. 
See \cite{mvp00}
for a perturbative discussion about a Wald \cite{wal74} field in Hawking's 
approach.
At the same time, the magnetosphere tends to equilibrate 
locally by charge-separation.
These two processes are generally in competition with one another, 
subject to current continuity linking asymptotic boundary conditions 
on the horizon and at infinity. In one proposal for the boundary
conditions, discussed below,
these considerations
indicate that outflows may be created by a macroscopic gap in
an open flux-tube
along the axis of rotation of the black hole.\cite{mvp01b,bro01}
\begin{figure}
\centerline{\epsfig{file=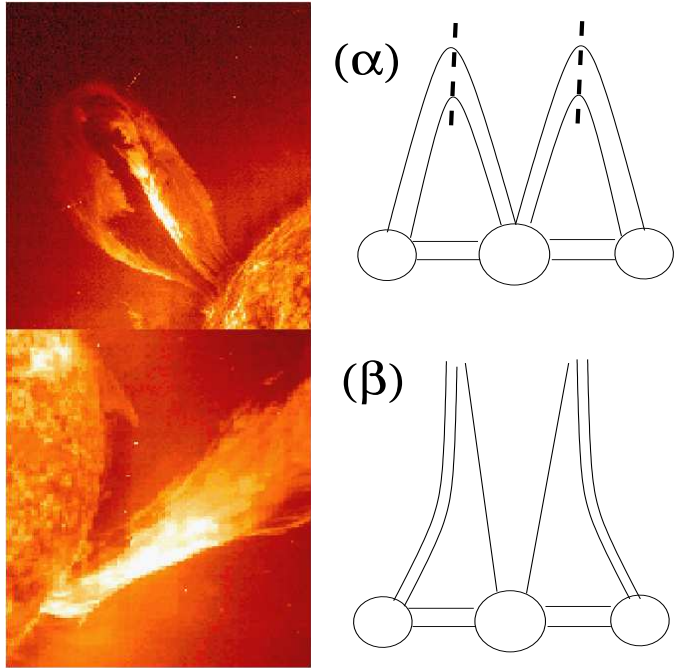,height=110mm,width=100mm}}
\caption{Images of closed ({\em top left}) and open ({\em bottom left})
         topology of flux-tubes in the solar atmosphere
		 from the Solar Heliospheric Observatory (Courtesy of
		 the SOHO/EIT consortium, ESA-NASA). 
		 The footpoints of the tubes are rooted in the
		 surface of the Sun with no-slip boundary conditions. The
		 open tube represents a violent  coronal mass ejection.
		 The present proposal builds on potentially similar 
		 structures in a torus magnetosphere around a black hole,
		 wherein the first corresponds to field-lines connecting the torus 
		 (with no-slip boundary conditions) to the black hole
		 (with slip boundary conditions) as schemetically
		 indicated in $(\alpha)$; 
		 the second corresponds to
		 field-lines extending from the black hole to infinity
		 (with slip boundary conditions on the horizon) and from the torus
		 to infinity (with no-slip boundary conditions on its surface) as
		 schemetically indicated in $(\beta)$. Since open flux-tubes ($\beta$)
		 form from closed loops $(\alpha)$, these fluxes are the same in
		 magnitude and opposite in sign for the black hole and the torus. 
		 This co-axial flux-structure permits current closure at infinity.}
\end{figure}

An open flux-tube supported by the black hole is endowed with
ingoing boundary conditions on the horizon of the 
black hole and outgoing boundary conditions at infinity.
In field theory, these asymptotic boundary conditions assume
conjugate radiative-radiative boundary conditions. In the
continuum limit of a plasma which is asymptotically in charge-separated
equilibrium,
these become slip-slip boundary conditions: the angular
velocities of the flux-surfaces on the horizon may differ from
that of the black hole, and the angular velocity at infinity may
be non-zero.
In contrast, a flux-surfaces supported by baryonic matter 
are fixed to its angular velocity, namely that of the disk or torus
(a no-slip boundary condition).
This will hold to within a fair approximation over 
an appreciable scale relative to the system size.
Recall that this well-known corotation law is based on the
singular limit of perfect conductivity; 
deviations of order unity will arise
over distance scales of order $1/\epsilon$, upon deviations from
the corotation charge-density to order $\epsilon$.

Equilibration towards a force-free state
introduces an asymptotic null-condition on the current carried by
the flow going into the black hole: $j^2\rightarrow0$ upon 
approaching the horizon, where $j^b$ denotes the four-current.\cite{pun90,mvp01b}
This expresses the condition that the current
becomes asymptotically convective: $j^r=\pm \alpha j^t$, where $\alpha$ denotes
the redshift factor on-axis of the Kerr black hole. In approaching
the horizon, drift-currents are suppressed by a divergent Lorentz factor.
Here, we shall consider the proposal that the boundary
condition at infinity is similar in an ultrarelativistic outflow.
It would be of interest to study this proposal 
self-consistently with the micro-physics in the gap.

Frame-dragging appears explicitly in the expression for the
electric charge-density $\rho$ in the equilibrium charge-separated limit.
Indeed, the equilibrium charge-density is associated with a time-like coordinate 
which is orthogonal to the azimuthal Killing vector. Hence, this charge-density
corresponds to the density-at-infinity as seen by
zero angular momentum observers (ZAMOs); the ``true" angular velocity
of a flux-surface is that relative to a local ZAMO with angular
velocity $-\beta$. Consequently,
we have the expression $\rho=-(\Omega+\beta)B/2\pi$ as a modified
Goldreich-Julian density.\cite{gol69,bes97,hir98,mvp01b}
The asymptotical condition
$j^2=0$ on the horizon and infinity now expresses electric current mediated by
convection of this modified Goldreich-Julian density.
Integrating over an effective area corresponding to a given flux
surface $A_\phi=$const., we find {\em current sources}
$I_-=\Omega_-A_\phi$ at infinity
and $I_+=(\Omega_H-\Omega_+)A_\phi$ on the horizon. Here,
$\Omega_-$ and $\Omega_+$ denote the Boyer-Lindquist angular velocities
of the two asymptotically equilibrated sections attached to infinity and the
horizon, respectively.
Current continuity enforces the condition $\Omega_-=\Omega_H-\Omega_+$.

Global current closure may obtain over the surrounding torus.
Here, we appeal to a potential similarity to solar flares, 
as observed by the Transient Region Corona Experiment
(TRACE) and the Solar Heliospheric Observatory (SOHO). While
magnetic field-lines form closed loops when supported by
compactly supported current sources, these loops are occasionally 
unstable, and flare as open prominences, as shown in the left 
column of Fig. 4.
This suggests that open flux-tubes might form from loops
in the torus magnetosphere connecting the black hole and the torus,
sketched in the right column of Fig. 4. 
If so, this gives rise to an inner open flux-tube supported
by the magnetic moment of the black hole and outer open flux-tube
supported by the magnetic moment of the inner face of the torus.
Creating open flux-tubes in the fashion is accompanied by an 
algebraic constraint: the inner and outer
flux-tubes carry a magnetic flux which is equal in magnitude and 
opposite in sign.
Applying the same asymptotic boundary condition $j^2=0$ to the
outflow from the torus -- and this is expected to be an approximation
to within order unity -- we obtain global current closure in the form
of $I_-=I_+=I_T=\Omega_TA_\phi$, where $\Omega_T$ denotes the
angular velocity of the torus.
The result is a differentially rotating gap between forementioned two
equilibrium sections, one attached to infinity and the other
attached to the horizon, with a 
Faraday-induced potential drop $\Delta V=(\Omega_+-\Omega_-)A_\phi$.
The power dissipated in this gap becomes
\begin{eqnarray}
P=\Omega_T(\Omega_H-2\Omega_T)A^2_\phi.
\end{eqnarray}
 Thus, a gap forms
with macroscopic dissipation whenever the black hole spins
more rapidly than {\em twice} the angular velocity of the torus.\cite{mvp01b}

Of some interest is the formation of the gap 
$(\Omega_H>2\Omega_T$) while being in a state of hyper-accretion
$(\Omega_H<\Omega_H^*$). Attributing the power released in the
gap to the input to the observed GRB and afterglow emissions,
suggests the possibility for afterglow emissions to short bursts. 
Unless the environment to short bursts is dramatically different
from long bursts, HETE-II should see afterglows to short bursts
as well.\cite{mvp01a}

\subsection{Clustering and spread in GRB emissions}

Recent analysis of achromatic breaks in a sub-sample of
GRB lightcurves indicates that these emissions
are beamed, and that their true fluence $E_{grb}$ is
standard 
(with a dynamic range of about one decade) at few times $10^{50}$ergs
\cite{fra01,pan01}.
At the same time, the opening angle displays a rather
wide dynamic range, between a few degrees and a few tens of degrees.

In \cite{mvp01c}, we consider a geometrically standard 
inner region in the vicinity of the black hole, when
the torus is thick relative to the size of the black hole.
In this event, the opening angle of the
open flux-tube on the horizon commensurate with the
true emissions in GRBs is about $35^o$. Collimation of this 
flux-tube down to an opening angle $\theta_j$
on the celestial sphere may derive from winds coming off
the torus, possibly so along forementioned outer flux-tube.
This will account for a true output in outflow of about 
$E_{grb}(\theta_H/\theta_j)\epsilon^{-1}$, 
where $\epsilon\sim 0.15$ denotes the efficiency of kinetic
energy to gamma-rays. Though substantial, this output
remains well below the energy deposit into the torus by the
black hole. This predicts a bound
\begin{eqnarray}
\theta_j\le 35^o
\end{eqnarray}
in any large sample of observations.
We attribute variations in the opening angle
$\theta_j$ to a diversity in torus parameters.

\section{GRBs: the tip of the iceberg?}

For long bursts, the equivalence in poloidal topology
to pulsar magnetospheres indicates a high incidence of
the black hole luminosity into the torus. Only a small
fraction of less than one percent of the black hole
output is associated with the true output in 
GRB-afterglow emissions.

\subsection{Gravitational radiation from a torus around a black hole}

The torus processes the input from the black hole by emission 
in various channels. This can be estimated in a suspended accretion state,
including gravitational radiation, neutrino emissions
and Poynting flux-winds. Gravitational radiation will
be emitted as the torus develops non-axisymmetries,
which features several aspects which suggest considering
long GRBs as potential sources for LIGO/VIRGO.
Namely, the torus is strongly coupled to the spin-energy
of the black hole; lumpiness in the torus will produce
gravitational radiation at twice the Keplerian angular
frequency, i.e., in the range of 1-2kHz; the emission
in gravitational radiation should dominate over emissions
in radio waves; the true rate of GRBs should be frequent
as inferred from their beaming factor of a few hundred.
These gravitational wave emissions from the
torus are powered by the spin energy of the black hole. This
sets it apart from such emissions in neutron star-neutron star
mergers or by fragmentation in collapse towards supernaovae.

Non-axisymmetries in the torus are expected from dynamical
and, potentially, radiative instabilities. Notably so,
a geometrically thick torus, consistent with the recent
indication that long GRBs may be standard, is generally
subject to the Papaloizou-Pringle instability \cite{pap84}. If the torus
reaches a mass on the order of that of the central black hole,
it will be unstable to self-gravity. Of interest is further
the possibility of a Chandrasekhar-Friedman-Schutz instability,
or radiative instabilities since lumps of matter radiate
preferentially on inner orbits. It would be of interest to
study these radiative instabilities in further detail.
The resulting gravitational wave-emissions may be
quasi-periodic (QPO). This may be reminiscent of the
observed QPOs in accretion disks in X-ray binaries, some
of which have been attributed to general relativistic
effects in orbital motions \cite{ste00}.

A detailed calculation in the suspended-accretion state
gives the estimate $L_{gw}\simeq L_H/3$ for the luminosity
in gravitational waves as a fraction of the black hole
luminosity $L_H$.\cite{mvp01d} This gives a fluence in gravitational
waves 
\begin{eqnarray}
E_{gw}\simeq 1\% M
\end{eqnarray}
in terms of the mass $M$ of a rapidly spinning Kerr
black hole. This suggests to consider
black hole-torus systems as potential LIGO/VIRGO
sources of gravitational waves.
The abundance of GRBs on a cosmological scale suggests to
consider an interesting contribution to the stochastic
background in gravitational radiation.\cite{aur01,cow01}

\subsection{Calorimetric evidence for Kerr black hole}

The existence of black hole remains circumstantial, in 
particular for the population of stellar mass black holes.
Obtaining evidence based on first principles is a challenging
task which occasionally drives new observational strategies.
Perhaps, then, LIGO/VIRGO may contribute by calorimetry on
gravitational wave emissions. Successful
detection of a burst in gravitational waves of a duration
commensurate with the redshift corrected duration of 10-15s
of long bursts would provide evidence for
a compact and high-angular momentum inner engine. 
Additionally, a fluence in gravitational waves in excess of
the rotational energy of a rapidly spinning neutron star
would support the presence of a central Kerr black hole.
Kerr black holes have the unique property of storing up
to about a third of their mass in rotational energy. This
has no baryonic counter part.

This may be pursued by the observable combination
\begin{eqnarray}
\alpha=2\pi E_{gw}f_{gw}.
\end{eqnarray}
Here, $\alpha$ is dimensionless upon use of geometrical units,
with energy expressed in terms of the corresponding Schwarzschild
radius and frequency in 1/cm. It may be noted that
$\alpha$ is a compactness parameter reminiscent of the dimensionless 
Kerr parameter $a/M$. Indeed, for large specific
angular momenta $a$, the ratio $a/M$ becomes effectively a measure for
the stored rotational energy to the linear size of the black hole.
The estimates above show that a black hole-torus system may produce 
$\alpha=0.01-0.015$, while a neutron star satisfies $\alpha<0.007$. 
This indicates an opportunity to detect $\alpha$ in excess of that
permitted by a neutron star, to serve as calorimetric evidence of a 
Kerr black hole.  Notice that $E_{gw}$ requires a distance
estimate to the source. In practice, this may require statistical
analysis on a sample of detections.
   
   {\bf Acknowledgment.}
   The author thanks the Korean Institute for Advanced Study for
   their hospitality and for hosting a very stimulating meeting.
   He also thanks G.E. Brown, C.W. Lee and A. Levinson for 
   continuing conversations. This work is partially supported by 
   NASA Grant No. 5-7012 and an MIT C.E. Reed Award.

\mbox{}\\

\end{document}